# Chaotic Analog-to-Information Conversion with Chaotic State Modulation

Sheng Yao Chen, Feng Xi and Zhong Liu
Department of Electronic Engineering
Nanjing University of Science & Technology
Nanjing, Jiangsu 210094
The People's Republic of China
E-mails: chen_shengyao@163.com, xf.njust@gmail.com, eezliu@njust.edu.cn

**Abstract:**

Chaotic compressive sensing is a nonlinear framework for compressive sensing. Along the framework, this paper proposes a chaotic analog-to-information converter, *chaotic modulation*, to acquire and reconstruct band-limited sparse analog signals at sub-Nyquist rate. In the chaotic modulation, the sparse signal is randomized through state modulation of continuous-time chaotic system and one state output is sampled as compressive measurements. The reconstruction is achieved through the estimation of the sparse coefficients with principle of chaotic impulsive synchronization and $l_p$-norm regularized nonlinear least squares. The concept of supreme local Lyapunov exponents (SLLE) is introduced to study the reconstructablity. It is found that the sparse signals are reconstructable, if the largest SLLE of the error dynamical system is negative. As examples, the Lorenz system and Liu system excited by the sparse multi-tone signals are taken to illustrate the principle and the performance.

# 1  Introduction

Shannon-Nyquist sampling theorem states that the sampling rate of traditional analog-to-digital conversion (ADC) is at least twice the signal bandwidth to attain alias-free sampling [1, 2]. The persistently increased demand for signals with higher bandwidth motivates to search innovative ADCs, particularly when sampling rates reach several gigahertzs. Recently developed compressive sensing (CS), also called compressive sampling [3, 4], bring us a new perspective on sub-Nyquist signal acquisition. In CS theory, sparse signals are sampled by random linear projection, and their sampling rates are close to their information rates (sparsity levels) rather than their bandwidths. Under proper restricted isometry property (RIP) condition [5], the sparse signals can be recovered correctly.

Along the CS theory of discrete-time sparse signals, several schemes have been proposed to implement the sub-Nyquist sampling of sparse analog signals. Among them are random sampling [6], random filtering [7], random demodulation [8], modulated wideband converter [9] and so on. Parallel to ADC, these schemes are named as analog-to-information conversion (AIC). Even though they are different in their implementation structures, the essence of all these schemes is to randomly measure the signals in time domain or frequency domain or time-frequency domain. Particularly, the random demodulation has become the basic unit of other complex AIC structures such as modulated wideband converter [9] and segmented compressed sampling [10].

It should be noted that the CS theory and its corresponding AICs are developed from linear random measurements. Recently, some extensions have been made to perform nonlinear compressive measurements [11-14]. In [14], we presented a chaotic CS (ChaCS), which implements the sub-Nyquist sampling by down-sampling state variable of chaotic systems excited by the sparse signals. In comparison with linear CS theory, the ChaCS has several advantages, such as simple implementation structure, security of measurement data, reproductivity of the measurement system at a remote agent. However, the complexity of signal reconstruction increases because of the nonlinearity of the measurement systems.

In this paper, we propose a chaos-based AIC for sub-Nyquist sampling of sparse analog signals with ChaCS theory. It will be clear that the proposed AIC is named as *chaotic modulation*. The chaotic modulation consists of measurement subsystem and reconstruction subsystem, as shown in Fig.1. The measurement subsystem performs the compressive measurements by down-sampling the system state of the continuous-time system excited by the sparse analog signals. The reconstruction subsystem reconstructs the sparse analog signals through impulsive synchronization in conjunction with parameter estimation. Similar to the random demodulation, the chaotic systems plays a role of randomizing the excitations. There are several ways to implement the excitation as in chaotic modulation communications [15]. These include additive state masking and multiplicative state modulation. In this study, we take the multiplicative state modulation, *i.e*., a system state multiplied by the sparse signals, to implement the excitation. The proposed chaotic modulation is developed from the ChaCS theory and it inherits the advantages and disadvantages of the

ChaCS.

An essential issue in the ChaCS or the proposed chaotic modulation is whether the sub-Nyquist samples contain information enough to reconstruct the sparse signals. In linear CS theory, this is well-developed. If the measurement matrix satisfies the RIP condition, the sparse signal can be exactly reconstructed [5]. Some extensions have been made to nonlinear measurements [11-13]. However, they do not apply to ChaCS or the chaotic modulation because of the high nonlinearity in chaotic systems. For the chaotic modulation, we develop a sufficient reconstructability condition with chaotic impulsive synchronization theory. As shown in Fig.1, the excitation is reconstructable if the reconstruction subsystem is synchronized with the measurement subsystem. Then the reconstructability can be developed from the synchronizability of the two subsystems. In [16], we introduced a concept of supreme local Lyapunov exponents (SLLEs) for describing the impulsive synchronization. With the chaotic modulation, it is found that if the largest SLLE of the error system between measurement subsystem and reconstruction subsystem is negative, the excitation is reconstructable. At the same time being, this condition directs us the selection of the sampling interval and the sampled state variable.

The remainder of this paper is organized as follows. In the next section, we introduce the signal model and problem. Section III describes the proposed chaotic modulation. In Section IV, Lyapunov criterion, a sufficient condition for the signal reconstructability is presented. In Section V, we analyze and simulate the reconstruction performance of chaotic modulation with Lorenz system and Liu system.

Finally, we give the conclusions in Section VI.

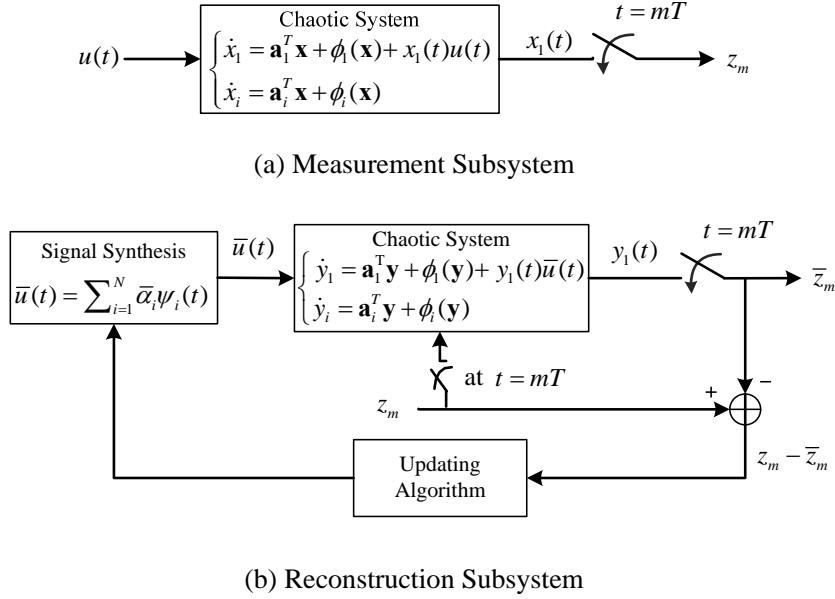

(a) Measurement Subsystem

(b) Reconstruction Subsystem

Fig.1 Chaotic Modulation Structure

## 2 Signal Model and Problem Statement

Consider a finite-time analog signal, $u(t)$ ($t \in [0, T_u]$), which can be expanded on the basis of $\{\psi_i(t)\}_{i=1}^N$.

$$u(t) = \psi(t)\boldsymbol{\alpha} = \sum_{n=1}^N \alpha_n \psi_n(t) \qquad t \in [0, T_u] \qquad (1)$$

where $\psi(t) = [\psi_1(t), \psi_2(t), \ldots, \psi_N(t)]$ is a basis vector and $\boldsymbol{\alpha} = [\alpha_1, \alpha_2, \ldots \alpha_N]^T$ is expanding coefficient vector. The signal $u(t)$ is called $K$-sparse if $\|\boldsymbol{\alpha}\|_0 = K \ll N$, where $\|\boldsymbol{\alpha}\|_0$ means the number of non-zero elements of $\boldsymbol{\alpha}$. Its sparsity is attributed to the basis.

This paper focuses on sparse multi-tone signals on real Fourier basis. A lot of signals in communication, acoustics and radar can be described by the model [8]. For the finite-time length $T_u$, its frequency resolution of Fourier basis is $\Delta f = 1/T_u$. For a band-limited signal with the bandwidth $W$, the signal is represented by

$$u(t) = \sum_{i=1}^{N/2} \alpha_i \cos(2\pi i \Delta ft) + \alpha_{i+N/2} \sin(2\pi i \Delta ft), \quad \text{for } t \in [0, T_u) \qquad (2)$$

where $N = 2WT_u$. To obtain alias-free samples of the signal, the sampling rate of ADC must be not less than $2W$ according to the Shannon-Nyquist sampling theorem. With random operation on the signal, CS theory samples the spare signal at its information rate which is far less than its Nyquist rate.

In this paper, we try to develop a randomizing structure with chaotic system and design a chaotic AIC structure with the ChaCS principle. What is more, we will establish a reconstructability condition of the sparse signal in virtue of chaotic impulsive synchronization theory.

## 3 Chaotic Modulation

Consider a continuous-time chaotic system

$$\dot{\mathbf{x}} = \mathbf{A}\mathbf{x} + \Phi(\mathbf{x}) \qquad (3)$$

where $\mathbf{x} \in \mathbb{R}^n$ denotes the state variables of system, $\mathbf{A} = [\mathbf{a}_1^T, \mathbf{a}_2^T, \cdots, \mathbf{a}_n^T]^T$ is an $n \times n$ constant matrix, $\mathbf{a}_i^T$ is the $i$ th row vector of $\mathbf{A}$, $\Phi = [\phi_1, \phi_2, \cdots, \phi_n]^T$ is a continuous nonlinear function vector satisfying Lipschitz condition $\|\Phi(\mathbf{x}) - \Phi(\hat{\mathbf{x}})\|_2 \leq L_1 \|\mathbf{x} - \hat{\mathbf{x}}\|_2$ with a Lipschitz constant $L_1$. In general, the system (3) will be chaotic for some range of the parameter matrix $\mathbf{A}$. Without loss of generality, we assume that the element $a_{11}$ of the matrix $\mathbf{A}$ is a flexible parameter and the system (3) is chaotic for a time-varying $a_{11}$ with its amplitude limited to the range. This characteristic is exploited to design the chaotic modulation structure.

*A. Measurement Subsystem*

Let the analog signal $u(t)$ be added to $a_{11}$. Then we have a non-autonomous chaotic system

$$\begin{cases} \dot{x}_1 = \mathbf{a}_1^T \mathbf{x} + \phi_1(\mathbf{x}) + x_1(t)u(t) \\ \dot{x}_i = \mathbf{a}_i^T \mathbf{x} + \phi_i(\mathbf{x}) \end{cases} \quad i = 2, \cdots, n \qquad (4)$$

where $u(t)$ is scaled so that the system (4) is chaotic. It is seen that the signal $u(t)$ is modulated on the system variable $x_1(t)$ to implement the excitation. With the modulation, the signal $u(t)$ is randomized by the state $x_1(t)$. We call the implementation as chaotic modulation. The bandwidth of $u(t)$ is assumed to be less than that of $x_1(t)$, otherwise the spectrum of $x_1(t)u(t)$ is out of the pass band of chaotic system and $u(t)$ cannot be sensed effectively.

Let the state variable $x_1$ be observable for simplicity. The measurements $z_m$ of $u(t)$ is defined as the period uniform sampling of $x_1$,

$$z_m = x_1(mT), \qquad m = 1, 2, \cdots, M \qquad (5)$$

where $T$ is sampling interval and $M$ is the number of measurements. For the signal of length $T_u$, $M = \lfloor T_u/T \rfloor$. When $M < 2WT_u$, the chaotic modulation is a sub-Nyquist sampling system of the analog signal. The measurement subsystem is illustrated in Fig.1 (a).

## B. Reconstruction Subsystem

Signal reconstruction is an essential ingredient in CS. Only under the signal be reconstructed properly, the chaotic modulation could be thought as an effective AIC structure. To perform the reconstruction of $u(t)$, or equivalently the estimation of $\boldsymbol{\alpha} = [\alpha_1, \alpha_2, ..., \alpha_N]^T$, from $z_m$, we construct a reconstruction subsystem with the chaotic impulsive synchronization theory [17,18]. Consider an impulsive system

$$\begin{cases} \dot{y}_1 = \mathbf{a}_1^T \mathbf{y} + \phi_1(\mathbf{y}) + y_1(t)\bar{u}(t), & t \neq mT \\ y_1 = x_1, & t = mT \\ \dot{y}_i = \mathbf{a}_i^T \mathbf{y} + \phi_i(\mathbf{y}), & i = 2,...n \end{cases} \qquad (6)$$

where $\bar{u}(t) = \sum_{i=1}^{N} \bar{\alpha}_i \psi_i(t)$, $\bar{\boldsymbol{\alpha}} = [\bar{\alpha}_1, \bar{\alpha}_2, ..., \bar{\alpha}_N]^T$ is the unknown coefficient vector. The system (6) is driven by the system (4) only at sampling instants. Then the system (4) and the system (6) are called the driving system and the driven system, respectively, in the framework of impulsive synchronization. Similar to (5), we can obtain

$$\bar{z}_m = y_1(mT), \qquad m = 1, 2, ..., M \qquad (7)$$

If $\bar{\boldsymbol{\alpha}} = \boldsymbol{\alpha}$ and $T$ is set properly, the system (6) can be synchronized with the system (4), i.e., $\mathbf{y}(t) \to \mathbf{x}(t)$ as $t \to \infty$. When $\bar{\boldsymbol{\alpha}}$ is unknown, the coefficient vector can be obtained through the synchronization-based parameter estimation for the synchronizable $T$. With $z_m$ and $\bar{z}_m$ available, the impulsive synchronization based reconstruction subsystem is constructed as shown in Fig.1 (b). The error outputs between $z_m$ and $\bar{z}_m$ are used to formulate the updating algorithm of $\bar{\boldsymbol{\alpha}}$. The updated $\bar{\boldsymbol{\alpha}}$ is then to synthesize the excitation signal. Upon synchronizing, $\bar{z}_m \to z_m$ and $\bar{\boldsymbol{\alpha}} \to \boldsymbol{\alpha}$.

With $z_m$ from the driving system and $\bar{z}_m$ from the driven system, the key problem is to design an updating rule for estimating $\bar{\boldsymbol{\alpha}}$ such that $\bar{z}_m = z_m$. The unknown $\bar{\boldsymbol{\alpha}}$ is implicitly included in $\bar{z}_m$. Let $\bar{z}_m = H_m(\bar{\boldsymbol{\alpha}})$ and $H_m(\bar{\boldsymbol{\alpha}})$ is determined by the reconstruction subsystem and $z_m$. An intuitive method to estimate $\boldsymbol{\alpha}$ is to solve a nonlinear least squares (NLS) problem

$$\min_{\bar{\boldsymbol{\alpha}}} \|\mathbf{H}(\bar{\boldsymbol{\alpha}}) - \mathbf{z}\|_2^2 \tag{8}$$

where $\mathbf{H} = [H_1, H_2, ..., H_M]^T$, $\mathbf{z} = [z_1, z_2, ..., z_M]^T$, and $\|\mathbf{z}\|_2$ denotes the Euclidean norm of $\mathbf{z}$. There are a lot of well-developed algorithms [19] to solve the NLS problem. However, if the number of measurements $M$ is less than $N$, the simple least-squares (8) may leads to over-fit. Considering the sparsity of signal coefficients, as in linear CS reconstruction, we can use the regularization techniques to enhance the sparsity. A direct regularization is to do

$$\min_{\bar{\boldsymbol{\alpha}}} \|\mathbf{H}(\bar{\boldsymbol{\alpha}}) - \mathbf{z}\|_2^2 + \mu \|\bar{\boldsymbol{\alpha}}\|_0 \tag{9}$$

where $\|\bar{\boldsymbol{\alpha}}\|_0$ is the term of $l_0$-norm regularization and $\mu > 0$ is the regularization factor. However, the $l_0$-norm $\|\bar{\boldsymbol{\alpha}}\|_0$ is not continuous and results in NP-hard problem. A proper relaxation is to solve the regularized NLS as

$$\min_{\bar{\boldsymbol{\alpha}}} \|\mathbf{H}(\bar{\boldsymbol{\alpha}}) - \mathbf{z}\|_2^2 + \mu \|\bar{\boldsymbol{\alpha}}\|_p^p \tag{10}$$

where $\|\bar{\boldsymbol{\alpha}}\|_p^p$ is $l_p$-norm of $\bar{\boldsymbol{\alpha}}$ ($0 \leq p \leq 1$). For the setup of the paper, we find that the regularization with $p = 0.5$ gives a reasonable estimate of $\boldsymbol{\alpha}$. As an extension of IRNLS algorithm [14], a $\varepsilon$-regularized IRNLS algorithm is proposed to solve (10). The algorithmic framework is given in appendix.

## 4 Lyapunov Criterion for Signal Reconstructability

As discussed in last section, the coefficient vector $\boldsymbol{\alpha}$ is able to estimate, or the sparse signal $u(t)$ is reconstructable, if the driven system is synchronized to the driving system for a proper setting of the sampling interval $T$. For CS applications, we expect large interval $T$, which implies the low sampling rate. Then under what conditions on $T$, can we ensure the system (4) synchronizable to the system (6) for a given system (3) and unknown sparse signal $u(t)$? In CS theory, this condition is called reconstruction condition. This section addresses the problem.

First, we consider the impulsive synchronization criterion when the sparse signal $u(t)$ is absent. There are several criteria [16-18] which can be used to study the synchronization. We find that the supreme local Lyapunov exponents (SLLEs) defined in [16] are most suitable for our problem setting. The SLLEs are the supremum of local Lyapunov exponents (LLEs) [20]. For systems (4) and (6) without the signal $u(t)$, the error system is given by

$$\begin{cases} \dot{e}_1 = \mathbf{a}_1^T \mathbf{e} + \phi_1(\mathbf{y}) - \phi_1(\mathbf{x}), & t \neq mT \\ e_1 = 0, & t = mT \\ \dot{e}_i = \mathbf{a}_i^T \mathbf{e} + \phi_i(\mathbf{y}) - \phi_i(\mathbf{x}), & i = 2,...n \end{cases} \quad (11)$$

where $\mathbf{e} = \mathbf{y} - \mathbf{x}$ is synchronization error. Then the LLEs are defined as

$$\lambda_i^L(\mathbf{x}_0, T_L) \equiv \frac{1}{T_L} \ln\left(\|\mathbf{H}_e(\mathbf{x}_0, T_L)\mathbf{v}_i(\mathbf{x}_0)\|\right) = \frac{\ln(\sigma_i(\mathbf{x}_0))}{T_L}, \quad i = 1,...n \quad (12)$$

where $\mathbf{x}_0$ is the initial state of driving system, $\mathbf{H}_e(\mathbf{x}_0, T_L)$ is a linear propagator which governs the time evolving of the error forward for a duration $T_L$, $\mathbf{v}_i(\mathbf{x}_0)$ is the $i$-th right singular vector of $\mathbf{H}_e(\mathbf{x}_0, T_L)$ and $\sigma_i(\mathbf{x}_0)$ is the $i$-th singular value. The SLLEs are given by

$$\lambda_i^S(T_L) \equiv \sup_{\mathbf{x}_0} \frac{1}{T_L} \ln\left(\|\mathbf{H}_e(\mathbf{x}_0, T_L)\mathbf{v}_i(\mathbf{x}_0)\|\right), \quad i = 1,...n \quad (13)$$

Different from LLEs, SLLEs do not depend on the initial states and are closely related to the duration $T_L$. In this sense, the SLLEs characterize the extreme expansion rate of the error system in finite duration in despite of the state locations on the synchronization manifold. It is shown that the origin of the error system is asymptotically stable, which means that the synchronization is asymptotically stable, if the largest SLLE is negative [16]. Furthermore, the condition allows large sampling interval over other synchronization criteria. However, for CS applications, we usually transmit sampling information of only a state variable to reduce the implementation complexity. In this case, we need to find a suitable state variable for compressive measurements.

When the unknown sparse signal $u(t)$ is present, the system (4) is a time-varying parameter system. The above-mentioned criterion is still applicable[1]. In general, the SLLEs will be different from those by normal parameter and the synchronizable sampling intervals will change. To ensure the change slight, we should select the parameter such that the modulation does not have much effect on the SLLEs. This is a parameter-dependent problem.

Now we take Lorenz system [21] as an example to illustrate the selection of the state variable and the effect on the SLLEs. The dimensionless Lorenz system is described as

$$\begin{cases} \dot{x}_1 = \sigma(x_2 - x_1) \\ \dot{x}_2 = rx_1 - x_2 - x_1 x_3 \\ \dot{x}_3 = x_1 x_2 - bx_3 \end{cases} \quad (14)$$

which is chaotic for $\sigma = 30$, $r = 50$ and $b = 3$.

As the sparse signal $u(t)$ is absent, Fig.2 illustrates the largest SLLE of the error system when the driven system is impulsively driven by different states of driving Lorenz system. In the simulation, the duration of SLLEs is selected as $T_L = 50$. It is seen that $x_2$ has the largest sampling interval for the impulsive synchronization among three state variables. Then we may take the samples of $x_2$ as the compressive measurements. It is also noticed that the largest SLLE does not show much changes for different parameters $r$ in chaotic ranges, as shown in Fig.3. Fig.4 shows the normalized magnitude spectrum of $x_1$. It can be calculated that the bandwidth of $x_1$ is about 5.15. (The bandwidth of chaotic signal is defined as the least bandwidth that concentrates 98% energy of the signal [22]). With these observations, we take the parameter $r$ as the flexible parameter. Then we can formulate the state modulated Lorenz system as

$$\begin{cases} \dot{x}_1 = \sigma(x_2 - x_1) \\ \dot{x}_2 = rx_1 - x_2 - x_1 x_3 + u(t)x_1 \\ \dot{x}_3 = x_1 x_2 - bx_3 \end{cases} \quad (15)$$

---

[1] When the sparse signal is present, the system (4) is non-autonomous and may be hyper-chaotic. However, the above-mentioned criterion is still applicable because the synchronization only requires the largest SLLE of the error system to be negative.

To sense the sparse signal efficiently, we limit the bandwidth of $u(t)$ to be less than 5.15. For chaotic motions, the maximal amplitude of $u(t)$ is confined to be less than 4. Assume that the bandwidth of $u(t)$ is $W = 5.0$ and the finite-time length of $u(t)$ is $T_u = 10$. Then the number of the basis $N$ is equal to $100$. Fig.5 shows the largest SLLE of the error system by the state modulated Lorenz system for different sparsity levels. It is seen that the largest SLLE changes slightly for different sparsity level, which implies that the sparsity level of the signal has little effect on the signal reconstructability. Moreover, it is found that the signal is reconstructable when $T \leq 0.25$, which will be used in the next section.

***Remark***: The reconstructability is derived from the synchronizability. Several synchronization criteria can be used to check the reconstructability. As discussed in [16], these criteria often allow small synchronization intervals and are not applicable to compressive sensing. The problem with Lyapunov criterion is that we cannot set the sampling interval explicitly. For the application, we have to determine the sampling intervals by numerical calculations.

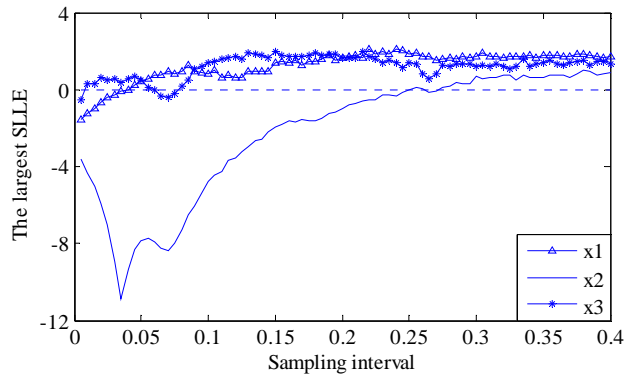

Fig.2 Variation of the largest SLLE as sampling intervals for different state variables

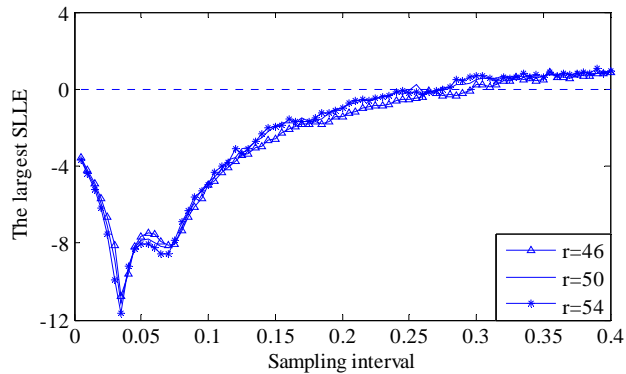

Fig.3 Variation of the largest SLLE as sampling intervals for different parameter $r$

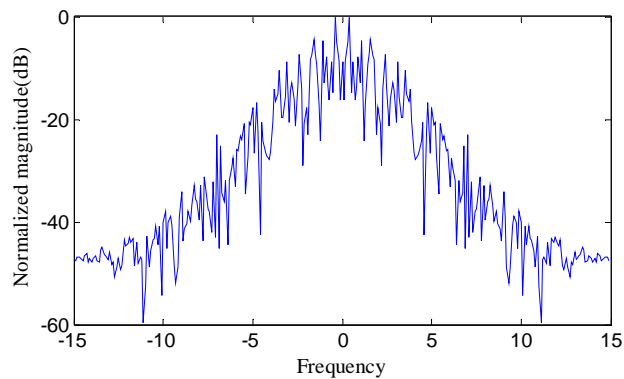

Fig.4 Normalized magnitude spectrum of $x_1$

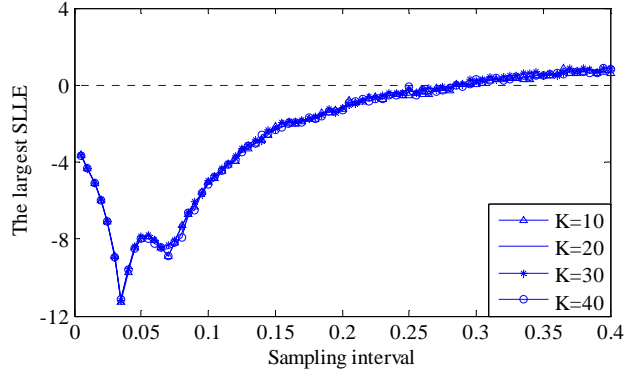

Fig.5 Variation of the largest SLLE as sampling intervals for different sparsity level signals

## 5 Simulation Experiments

In this section, several simulations are implemented to demonstrate the reconstruction performance of chaotic modulation. Lorenz system and Liu system [23] are exploited to construct chaotic modulation, respectively. Both of them have large synchronizable sampling intervals.

The dimensionless Liu system is described as

$$\begin{cases} \dot{x}_1 = \sigma(x_2 - x_1) \\ \dot{x}_2 = rx_1 - x_1 x_3 \\ \dot{x}_3 = cx_1^2 - bx_3 \end{cases} \quad (16)$$

which is chaotic for $\sigma = 30$, $r = 42$, $b = 2.5$ and $c = 4$. For chaotic modulation, the parameter $r$ is taken as a flexible one and the sparse signal $u(t)$ is modulated on the state $x_1$. For the parameter setting, the bandwidth of $x_1$ is 5.60. Then the highest frequency of the signal $u(t)$ is restricted to be less than 5.60. For the state variable $x_2$, doing the analysis similar to the last section, we find that the Liu system is synchronizable for the sampling interval $T \leq 0.25$. Then the samples of the state variable $x_2$ are adopted as compressed measurements.

The simulation scenarios are set up as follows: Sparse multi-tone signals are generated by real Fourier bases. The maximal bandwidth of the sparse signals is set as $W = 5.0$. The finite-time length of the signals is assumed to be $T_u = 10$. Then the frequency interval is $\Delta f = 0.10$ and the number of the basis is $N = 100$. The Nyquist sampling interval for the assumed signals is 0.1. The sparse positions are uniformly distributed over frequency range $[0, 5.0]$, and the sparse coefficients are from two types of distributions: *i.i.d.* Gaussian with zero mean and unity variance and *i.i.d.* Bernoulli with entries $\pm 1$. Different sparsity levels and sampling intervals are considered. The exciting signal $u(t)$ is scaled in order to guarantee the modulated systems are chaotic. A fourth-order Runge-Kutta method with step size $10^{-3}$ is used to solve ordinary differential equation.

The $\varepsilon$-regularized IRNLS algorithm in the appendix is used to estimate the sparse coefficients and $l_{0.5}$-norm is exploited to regularize the sparse coefficients. The initial setting of each coefficient for $\varepsilon$-regularized IRNLS is randomly set over interval $[-1,1]$. Other parameters $\mu$, $\varepsilon_0$, $c$ and $\lambda$ are set to be $10^{-2}$, $10^{-2}$, $10^{-1}$ and $10^{-1}$, respectively. The $\varepsilon$-regularized IRNLS is deemed convergence if the relative error (stopping criterion) between two consecutive iterations is less than $10^{-3}$. The median relative error of estimated coefficients is used to measure reconstruction performance of the sparse signal,

$$Err = \|\bar{\boldsymbol{\alpha}} - \overline{\boldsymbol{\alpha}}\|_2 / \|\boldsymbol{\alpha}\|_2$$

In the first realizations, the multi-tone signals with two sparsity levels ($K = 10$ and $K = 18$) are simulated with Lorenz system. The sampling interval is set as $T = 0.2$, *i.e.*, a half of Nyquist rate. Figs. 6 and 7 show one realization of chaotic modulation for the case of the Gaussian coefficients. Sparsity positions and amplitudes are shown in Fig.6 (a) and Fig.7 (a), respectively. It is seen from Fig.6 (a) that for $K=10$, the reconstructed signal matches the original one well and the sparse positions/coefficients are estimated correctly. The median relative error of the estimated coefficients is about $7.33\times10^{-3}$. For this case, it is deemed that the global solution is found and the excitation signal is accurately estimated. While for $K=18$, the estimated relative error is about $5.63\times10^{-2}$ and only is a local solution found. Nevertheless, the sparse positions are well estimated, as seen from Fig.7 (a). This is very important for the application of spectrum estimation. In addition, the reconstructed signal is still a reasonable approximation to the original one (Fig.7 (b)). From the point of view of waveform reconstruction, the sparse signal is well reconstructed, although the local optimal coefficients are found. The experiments illustrate that the chaotic modulation is able to implement the sub-Nyquist sampling for the selected examples. We also do simulations by the Liu system. The same conclusions can be drawn.

Next, we simulate the reconstruction performance for different sparsity levels and different sampling intervals and make comparisons between Lorenz system and Liu system. 100 realizations are conducted and the averaged results are shown in Fig.8 and Fig.9 with the two systems for two coefficient distributions. As seen from Fig.8 and Fig.9, the estimation performance decreases as the sparsity level increases. The reason is that the effect of the sparse regularization in (10) becomes poor for large sparsity levels and the problem (10) is easily trapped into local optimal solutions. Also noticed, the sampling intervals have the same effects as the sparsity. The larger the sampling intervals are, the fewer the measurements are, and the more the feasible solutions of (10) are. Then the probability of the local solutions for (10) increases. As for two coefficient distributions, the reconstruction performance of the Gaussian distribution is superior to that of Bernoulli one. The performance discrepancy comes from the IRNLS algorithm itself [14].

For the two systems, we find that the chaotic modulation with Lorenz system has better reconstruction performance than that with Liu system. This may be due to the fact that the cost function in (8) formulated from Lorenz system is more sensitive to the sparse coefficients than that from Liu system. Then the regularization term of Lorenz system has less effect on the global optimal solution of the problem (8) than that of Liu system.

Finally, we make a comparison between random demodulation and chaotic modulation. Fig.10 shows the median relative error of the random demodulation with different sampling intervals and sparsity levels. In the simulation, the chipping sequence switches between the levels $\pm 1$ randomly at the Nyquist rate 10Hz, and the $\varepsilon$-regularized IRLS algorithm with $p = 0.5$ [24] is used to perform the reconstruction. It is seen that the sparsity levels and the sampling intervals have the same effects on the reconstruction performance as the chaotic modulation. As the increase of the sparsity levels and/or the sampling intervals, the reconstruction performance of the random demodulation degrades. This is an inherent problem of the AIC systems. For high sparsity levels and/or large sampling intervals, the reconstruction errors will be large because the RIP condition is satisfied with less probability. However, the reconstruction condition of the proposed modulation is largely determined by the sampling interval. For the synchronizable systems, the degration of the reconstruction performance is due to the local solutions of (10). Simulations show that the proposed chaotic modulation with both Lorenz system and Liu system is in general superior to the random demodulation.

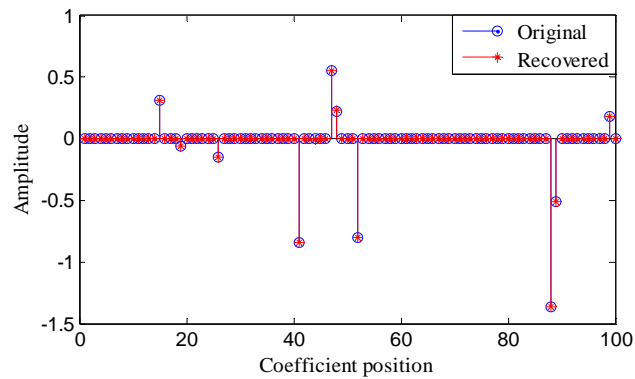

(a) Sparse coefficients and its estimation

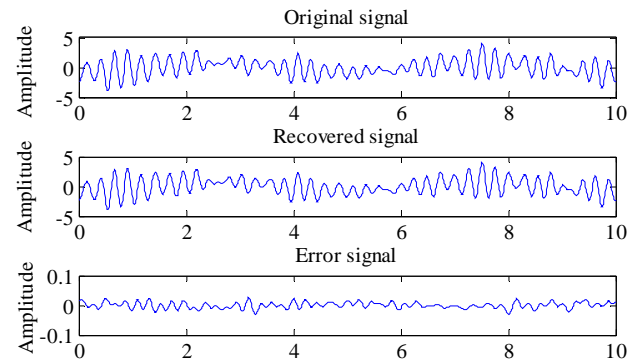

(b) Original waveform and its reconstruction

Fig.6 Sparse signal and its reconstruction for chaotic modulation with $K=10$

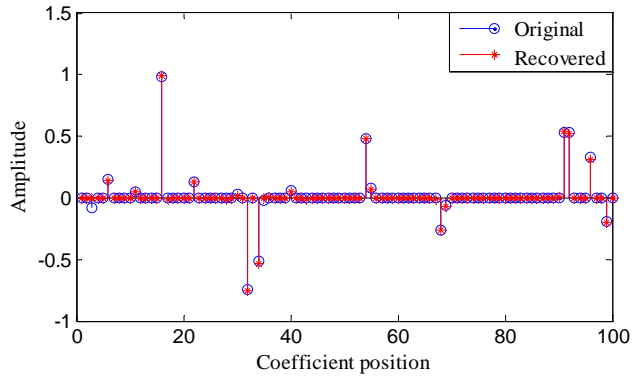

(a) Sparse coefficients and its estimation

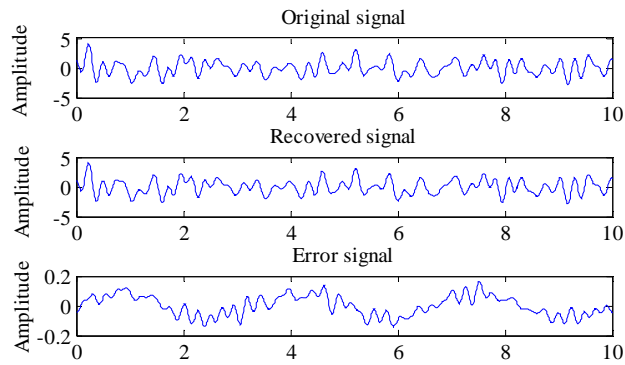

(b) Original waveform and its reconstruction

Fig.7 Sparse signal and its reconstruction for chaotic modulation with $K=18$

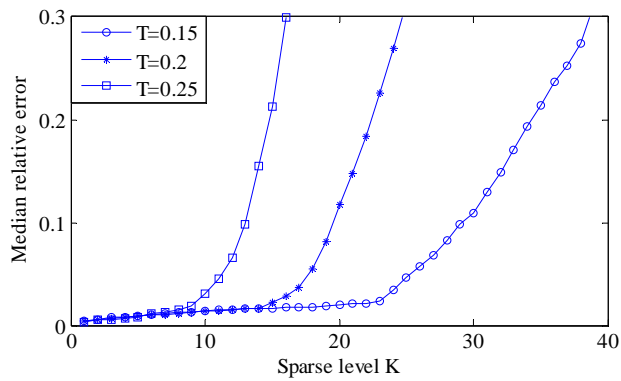

(a) Gaussian coefficients

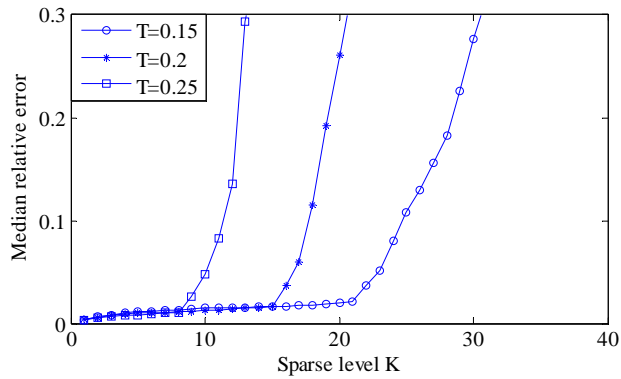

(b) Bernoulli coefficients

Fig.8 Median relative errors vs. sparsity for chaotic modulation with Lorenz system

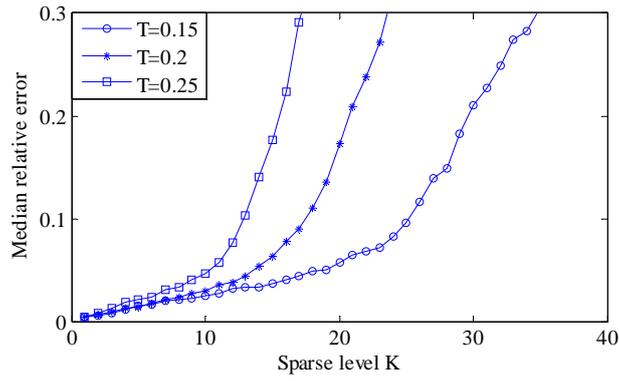

(a) Gaussian coefficients

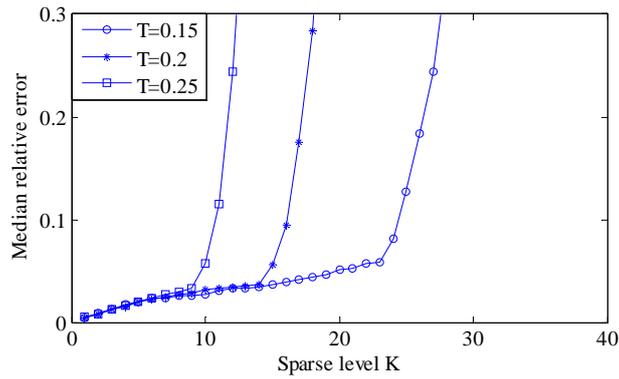

(b) Bernoulli coefficients

Fig.9 Median relative errors vs. sparsity for chaotic modulation with Liu system

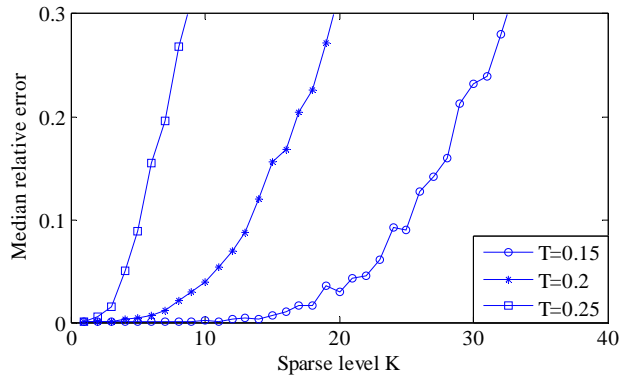

(a) Gaussian coefficients

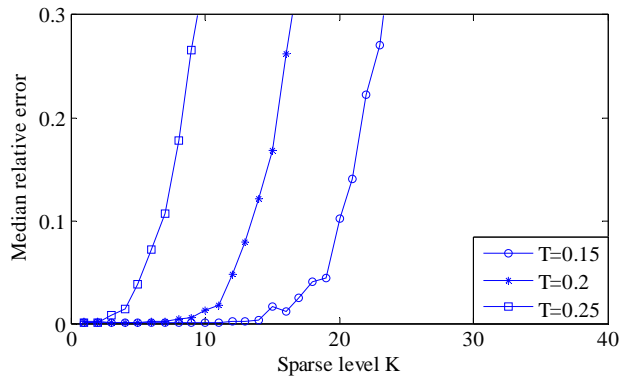

(b) Bernoulli coefficients

Fig.10 Median relative errors vs. sparsity for random demodulation

## 6 Conclusions

This paper has developed a chaotic modulation to implement the compressive sampling of sparse analog signals. The measurement and reconstruction principle is described. The reconstructability is studied with the concept of the SLLEs. Simulation results validate that the proposed chaotic modulation is an efficient AIC structure. Varieties of chaotic systems provide much freedom for the implementation.

The chaotic modulation is mainly built upon the chaotic impulsive synchronization theory. The reconstructability is then derived from the synchronizability. The analysis is different from the conventional CS, which is usually performed through random matrix theory. In comparison with random demodulation, the proposed chaotic modulation shows superior reconstruction performance.

In the chaotic modulation, the design of chaotic systems is important for practical applications. As discussed in last two sections, for a given sparse signal, the sub-Nyquist sampling rate is determined by the synchronizable sampling interval, while the synchronizable interval is a system-dependent problem. The design of the chaotic systems with large synchronizable interval is under investigation.

## 7 Acknowledgment

This work was supported by the National Science Foundation of China (60971090, 61171166, 61101193).

# 9  Appendix: $\varepsilon$-Regularized IRNLS Algorithm

In this appendix, we generalize the IRNLS algorithm [14] to solve (10). The fundamental idea is to replace the $l_p$-regularized term in (10) by a weighted $l_2$ norm with the objective function as,

$$\min_{\bar{\boldsymbol{\alpha}}} \left\| \mathbf{H}(\bar{\boldsymbol{\alpha}}) - \mathbf{z} \right\|_2^2 + \mu \left\| \mathbf{W}^{1/2} \bar{\boldsymbol{\alpha}} \right\|_2^2 \tag{A1}$$

where $\left\| \mathbf{W}^{1/2} \bar{\boldsymbol{\alpha}} \right\|_2^2$ is a first-order approximation of $\left\| \bar{\boldsymbol{\alpha}} \right\|_p^p$ with diagonal weighting matrix $\mathbf{W}$. The diagonal element is computed from the previous iteration $\bar{\boldsymbol{\alpha}}^{j-1}$ by $w_k^j = ((\bar{\alpha}_k^{j-1})^2 + \varepsilon)^{(p-2)/2}$ $(k = 1, 2, \cdots, N)$ ($\varepsilon$ is a small positive number in case of some $\bar{\alpha}_k^{j-1} = 0$). In implementation, the small number $\varepsilon$ is replaced by a monotonically non-increasing sequence $\varepsilon_j$ to improve reconstruction performance for $p < 1$ [24]. The algorithmic framework is described as follows:

$\varepsilon$-Regularized IRNLS Algorithm

**Objective**: Minimize (10) with respect to $\bar{\boldsymbol{\alpha}}$

**Input parameters**: $\mu, \varepsilon_0, c, \lambda, err$ (stopping criterion).

**Initialization**: Set $\bar{\boldsymbol{\alpha}}^0$ to be an arbitrary random number chosen over interval [-1, 1].

**Loop**: Set $j = 0$ and do

1. For $k = 1,...,N$, compute $w_k^{j+1} = ((\bar{\alpha}_k^j)^2 + \varepsilon_j)^{(p-2)/2}$.

2. Find $\bar{\boldsymbol{\alpha}}^{j+1}$ that minimizes $\|\mathbf{H}(\bar{\boldsymbol{\alpha}}) - \mathbf{z}\|_2^2 + \mu \|\mathbf{W}^{1/2}\bar{\boldsymbol{\alpha}}\|_2^2$.

3. If $\|\bar{\boldsymbol{\alpha}}^{j+1} - \bar{\boldsymbol{\alpha}}^j\|_2 / \|\bar{\boldsymbol{\alpha}}^j\|_2 \leq c\sqrt{\varepsilon_j}$, $\varepsilon_{j+1} = \lambda \varepsilon_j$, else $\varepsilon_{j+1} = \varepsilon_j$.

4. If $\|\bar{\boldsymbol{\alpha}}^{j+1} - \bar{\boldsymbol{\alpha}}^j\|_2 / \|\bar{\boldsymbol{\alpha}}^j\|_2 \leq err$, exit.

5. Set $j = j + 1$.

**Result**: Output the estimated $\bar{\boldsymbol{\alpha}}^{j+1}$.

The input parameter $c$ is a small positive constant and $\lambda < 1$ is an attenuation factor of $\varepsilon$ sequence. Minimization of $\|\mathbf{H}(\bar{\boldsymbol{\alpha}}) - \mathbf{z}\|_2^2 + \mu \|\mathbf{W}^{1/2}\bar{\boldsymbol{\alpha}}\|_2^2$ is a NLS problem. In the implementation, we use the *lsqnonlin* function in MATLAB Optimization Toolbox to find its solution.